\documentclass[aps,prl,10pt,reprint,superscriptaddress]{revtex4-2}

\usepackage{amsmath,amsfonts,amssymb}
\usepackage{xcolor,graphicx}
\usepackage{braket}
\usepackage{csquotes}
\usepackage[pdftex]{hyperref}
\hypersetup{
    colorlinks = true,
    linkcolor = blue,
    anchorcolor = blue,
    citecolor = red,
    filecolor = blue,
    urlcolor = blue}

\begin{document}

\title{Syncopated {B}essel beams}

\author{Irán Ramos-Prieto, Ulises Ruíz, Israel Julián-Macías, Francisco Soto-Eguibar, David {Sánchez}-{de-la-Llave}, Héctor M. Moya-Cessa}

\affiliation{Instituto Nacional de Astrofísica Óptica y Electrónica (INAOE)\\ Luis Enrique Erro 1, Santa María Tonantzintla, Puebla, 72840, Mexico}

\begin{abstract}
We introduce the \textit{syncopated Bessel beam}, a new class of exact solutions to the paraxial equation obtained by means of a sinusoidal modulation of the azimuthal phase at the source. This modulation imposes a phase rhythm that deliberately breaks the azimuthal symmetry, analogous to musical syncopation, and triggers a topological transformation that deflects the propagation trajectory and shifts the beam’s center of symmetry off the optical axis, while preserving its self-scaling invariance that can be explained by the Madelung-Bohm formalism. An exact analytical framework, supported by experimental validation, reveals the intrinsic structural robustness and preservation of topological properties through propagation. 
\end{abstract}
\maketitle

The introduction of self-scaling optical fields by Durnin~\cite{durnin1987} catalyzed a fundamental shift in the mastery of structured light, introducing beams with self-healing properties and invariant transverse topologies~\cite{bouchal2003,mcgloin2005,roadmap2017}. The canonical Bessel mode, synthesized from an infinitesimally thin annular spectrum, possesses an inherent cylindrical symmetry governed by a strictly linear azimuthal phase. While mathematically elegant, this uniform phase progression fundamentally restricts the beam's versatility in applications requiring asymmetric intensity distributions. To turn this ideal infinite-energy model into something that can actually be used, the Bessel-Gauss beam was introduced~\cite{gori1987}. By adding a Gaussian shape, the beam's energy stays finite while its center keeps its shape without spreading over a certain distance. Subsequent attempts to break this symmetry have led to substantial progress, most notably with the discovery of Airy beams~\cite{berry1979} and self-accelerating fields~\cite{siviloglou2007}. Although asymmetric Bessel-Gauss variants have been described as simple coordinate translations~\cite{kotlyar2014}, a unifying physical framework that relates angular phase gradients to spatial displacement remains elusive. Specifically, the bridge between a continuous azimuthal modulation and a discrete radial shift has yet to be fully exploited as an independent degree of freedom.

To overcome this limitation, in this letter, we introduce the \emph{syncopated Bessel beam}: a reconfiguration of the paraxial field manifold where the azimuthal symmetry is broken by a prescribed sinusoidal phase rhythm. In canonical Bessel modes, the azimuthal phase progresses linearly to maintain symmetric intensity profiles. By introducing a sinusoidal modulation, we establish a spatial syncopation---the nomenclature originating from the deliberate rhythmic disruption of the canonical azimuthal flow---where the phase gradient locally accelerates and decelerates along the annular source. This off-beat redistribution of the azimuthal rhythm manifests physically as a transverse phase-gradient pressure, which shifts the beam's center of symmetry off the axis while preserving its self-scaling invariance. Using Graf's addition theorem~\cite{abramowitz1964handbook}, we derive an exact analytical solution that justifies this syncopated formalism and provides an intuitive mapping between phase rhythm and spatial displacement. This mechanism offers a purely electromagnetic strategy for precise beam control, entirely bypassing the need for mechanical translation, and establishes a new paradigm in the manipulation of structured beams, where electromagnetic control circumvents the aberrations and extraneous parameters typically introduced by mechanical translation, thereby opening new avenues for topological control and advanced optical metrology.

The mathematical framework for syncopated Bessel beams is established by describing the propagation of an annular source modulated by a sinusoidal azimuthal phase rhythm. We consider a scalar monochromatic field $E(r,\theta,z)$ generated in the paraxial regime at the $z=0$ plane. The source is characterized by an infinitesimally thin annulus of radius $r_0$, modeled by the radial distribution $R(\rho) =R_0 \delta(\rho - r_0)$, where $R_0$ is a constant with correct dimensions, modulated by a syncopated phase profile $\Theta(\phi)=\exp[i(u\cos\phi - m\phi)]$; here, $u$ represents the modulation depth and $m$ denotes the topological charge. Under the Fresnel approximation, the field at a distance $z$ is governed by the diffraction integral, 
\begin{equation}
\begin{split}
E(r,\theta,z) =&-\frac{ik}{2\pi z} e^{\frac{ik r^2}{2z}} \int\limits_{0}^{\infty}\int\limits_0^{2\pi} \Theta(\phi) R(\rho) e^{\frac{ik\rho^2}{2z}}\\&\times e^{-\frac{ik r \rho\cos(\theta-\phi)}{z}} \rho \,d\rho \,d\phi,\\
\end{split}
\end{equation}
where $k=2\pi/\lambda$ is the wavenumber and which, using the Jacobi-Anger identity \cite{Colton1998,Cuyt2008}, can be written as
\begin{equation}
\begin{split}
E(r,\theta,z)=& -\frac{ik}{2\pi z} e^{\frac{ik r^2}{2z}}\sum_{n=-\infty}^{\infty}e^{in(\theta-\frac{\pi}{2})}\int\limits_0^{2\pi} \Theta(\phi)e^{-in\phi}d\phi \\&\times\int\limits_{0}^{\infty} R(\rho) e^{\frac{ik\rho^2}{2z}}J_n\left(\frac{kr\rho}{z}\right)\rho \,d\rho.
\end{split}
\end{equation}
Taking into account that $R(\rho) = R_0 \delta(\rho - r_0)$ and $\Theta(\phi) = \exp[i(u\cos\phi - m\phi)]$, the integrals can be evaluated, yielding
\begin{equation}
\begin{split}
E(r,\theta,z)=& -\frac{iR_0kr_0}{z} e^{\frac{ik(r^2+r_0^2)}{2z}}e^{\frac{im\pi}{2}}\\
&\times\sum_{n=-\infty}^{\infty}e^{in\theta} J_{n+m}(u)J_n\left(\frac{krr_0}{z}\right).
\end{split}
\end{equation}
This transformation from the diffraction integral to a discrete summation reveals an underlying algebraic structure common to all fields with cylindrical symmetry that preserves their transverse configuration. The process involves two fundamental steps: first, the expansion of the propagation kernel $e^{-i Z \cos(\theta-\phi)}$ with $Z = kr\rho/z$ using the Jacobi-Anger identity, which decomposes the field into an angular spectrum of Bessel modes. Second, the azimuthal integration of the modulated source $\Theta(\phi) = \exp[i(u\cos\phi - m\phi)]$ yields the coefficients $J_{n+m}(u)$, which weight each mode $n$ according to the modulation depth $u$. 

\begin{figure}
\centering\includegraphics[width=\linewidth]{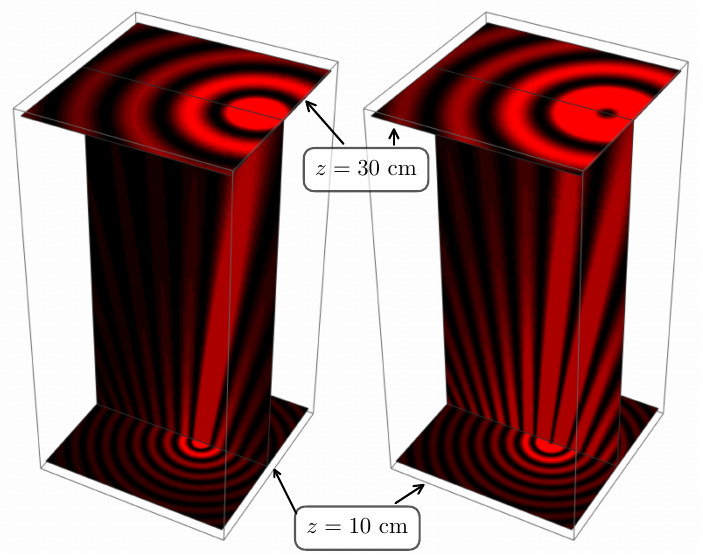}
    \caption{Three-dimensional volume rendering of syncopated Bessel beams for $m=0$ (left) and $m=1$ (right), generated using the same experimental parameters as in Fig.~\ref{fig_3}: He-Ne laser ($\lambda=633$ nm), annular source radius $r_0=120$ $\mu$m, and modulation depth $u=4$. The transversal observation window is $2 \times 2$ mm, which is essentially the size of the beam. These visualizations demonstrate the self-scaling propagation through space, where the transverse intensity distribution remains topologically invariant while expanding linearly with distance. The off-axis displacement of the beam's center of symmetry, governed by the modulation depth $u$, illustrates that the spatial syncopation acts as a robust mechanism for radial shift without compromising the beam's self-scaling nature.}
    \label{fig_1}
\end{figure}
Crucially, this infinite series of Bessel functions serves as a formal algebraic representation of a coordinate displacement mechanism within the paraxial framework. While canonical Bessel beams are traditionally defined as exact solutions to the cylindrical Helmholtz equation, the syncopated fields presented here emerge as a robust class of structured light that rigorously satisfies the paraxial wave equation despite the deliberate breaking of azimuthal symmetry. Physically, the sinusoidal phase rhythm $u\cos\phi$ induces a syncopation of the azimuthal flow, where the local acceleration of the phase gradient acts as a transverse phase gradient pressure. This spatial syncopation—a rhythmic disruption of the azimuthal symmetry—triggers a radial shift of the center of symmetry while establishing an off-axis trajectory that preserves the field's self-scaling invariance as an optical propagation mode within the paraxial regime. To obtain an exact closed-form solution that reveals this coordinate transformation, we apply the Graf addition theorem \cite[9.1.74, page 363]{abramowitz1964handbook}\cite[10.23 (i), page 246]{Olver2010}\cite[Eq. 1.13]{Dattoli1990}:
\begin{equation}
\sum_{n=-\infty}^\infty  t^n J_n(v)J_{n+m}(u) = f^m(v,u;t)\, J_m\!\big(g(v,u;t)\big),
\end{equation}
where the auxiliary functions are defined as $g(v,u;t) = \sqrt{v^2 + u^2 - vu(t + t^{-1})}$ and $f(v,u;t) = (u - v t^{-1})/g(v,u;t)$, with the parameter $t = \exp(i\theta)$ capturing the azimuthal dependence; this identity provides the necessary link to collapse the infinite series into a compact spatial field structure. Identifying the transverse coordinate $v = (k r_0 / z) r$ and specifying $t = \exp(i\theta)$ as the azimuthal phase factor, we apply the Graf identity to obtain the exact analytical expression for the electric field of the syncopated Bessel beam:
\begin{equation}\label{Central}
E(r,\theta,z) = -\frac{ i^{m+1}R_0kr_0}{z} e^{\frac{ik(r^2+r_0^2)}{2z}}\left( \frac{u - v e^{-i\theta}}{w} \right)^m J_m(w).
\end{equation}
The introduction of the syncopated coordinate $w = \sqrt{u^2 + v^2 - 2vu \cos\theta}$ represents a formal coordinate transformation that redefines the beam's symmetry axis. Moving beyond the static translation paradigm, the sinusoidal phase rhythm reconfigures the wave-front geometry, establishing a direct analytical link between angular topology and spatial displacement. Crucially, as the modulation depth vanishes ($u \to 0$), $w$ reduces to the conventional radial argument $v$, yet a fundamental distinction remains: the field recovered in this limit is not the non-diffracting Bessel beam of the Helmholtz equation~\cite{durnin1987}, but rather a paraxial Bessel mode—its structural analog within the paraxial wave equation. This mechanism employs a transverse phase-gradient pressure to guide the intensity distribution and its associated phase singularities along a stable, off-axis propagation trajectory. The existence of such an exact closed-form solution demonstrates that the transverse field structure can be shifted while maintaining its self-scaling invariance, enabling the all-electromagnetic steering of optical singularities with full topological stability. These features are illustrated in Fig.~\ref{fig_1}.

To reveal the physical mechanism underlying this topological stability, we examine the dynamical resilience of the syncopated field through the Bohm potential~\cite{Madelung_1927,Bohm_1952,Hojman2021,CRB}, $Q_{\texttt{Bohm}} = - (2k)^{-1} (\nabla_\perp^2 A / A)$. This formulation represents the internal potential structure that governs the diffractive dynamics, strictly guiding the transverse energy flow to sustain the field's self-scaling invariance. By substituting the amplitude of the central solution Eq.~\eqref{Central}, $A \propto J_m(w)$ and applying the corresponding geometric identities for the syncopated coordinate $w$, we obtain the exact analytical form of this guiding potential:
\begin{equation}\label{Potencial_Q}
Q_{\texttt{Bohm}}= \frac{1}{2}\left(\frac{r_0}{z}\right)^2\left(1 - \frac{m^2}{w^2}\right).
\end{equation}
What is perhaps most striking about Eq.~\eqref{Potencial_Q} is its profound algebraic simplicity. The entire structural complexity of the off-axis shift---driven by the deliberate rhythmic disruption of the azimuthal symmetry and the induced topological reconfiguration---is rigorously absorbed by the syncopated coordinate $w$. Rather than yielding a perturbed or fragmented potential, the governing dynamics emerge with absolute mathematical purity. This proves unequivocally that the syncopated field is not a mere distorted superposition but a fundamental, structurally pristine mode of propagation. The exactness of this potential dictates the beam's remarkable capacity to harbor a topological charge $m$ within a highly asymmetric intensity distribution, strictly sustaining the dynamical equilibrium required for self-scaling invariance.

\begin{figure}
\centering\includegraphics[width=\linewidth]{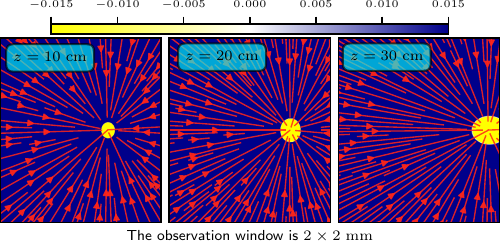}
    \caption{Transverse structural topology of the syncopated Bohm potential $Q_{\texttt{Bohm}}$ (color map) and its associated structural force field $\mathbf{F}_{\texttt{Bohm}} = -\nabla_{\perp} Q_{\texttt{Bohm}}$ (red arrows) for $m=1$, calculated using the same experimental parameters as in Fig.~\ref{fig_3}: He-Ne laser ($\lambda=633$ nm), annular source radius $r_0=120$ $\mu$m, and modulation depth $u=4$. The observation window is $2 \times 2$ mm, which is essentially the size of the beam. The visualization reveals an asymmetric potential well that acts as a stable structural guide. The converging force field traps the phase singularity at an off-axis equilibrium, ensuring that the self-scaling frame remains invariant despite the broken axial symmetry. The emergence of such a localized sink and the surrounding force vectors confirms that the syncopated modulation dynamically reconstructs the internal guiding manifold of the field.}
    \label{fig_2}
\end{figure}
As illustrated in Fig.~\ref{fig_2}, this exact potential distribution serves as the physical guide for the transverse energy flow. The propagation is dictated by the effective structural force $\mathbf{F}_{\texttt{Bohm}} = -\nabla_{\perp} Q_{\texttt{Bohm}}$, which governs the diffractive dynamics to maintain the self-scaling invariance of the field. In canonical Bessel beams, this potential exhibits azimuthal symmetry, yielding a purely radial force field. The sinusoidal phase rhythm deliberately disrupts this symmetry, exerting a transverse phase-gradient pressure that reconfigures the internal potential structure. This topological reconfiguration establishes an off-axis potential minimum that guides the phase singularity, securing the beam as a structurally pristine propagation mode. Hydrodynamically, the force maintains the highly asymmetric intensity distribution and rigorously preserves the topological charge throughout propagation. For $m=0$, the potential is evaluated as a globally shifted constant $Q_{\texttt{Bohm}}=2^{-1}(r_0/z)^2$; for $m \neq 0$, the phase rhythm reconstructs the guiding potential into an asymmetric well that robustly supports the displaced vortex core.

\begin{figure}
\centering\includegraphics[width=\linewidth]{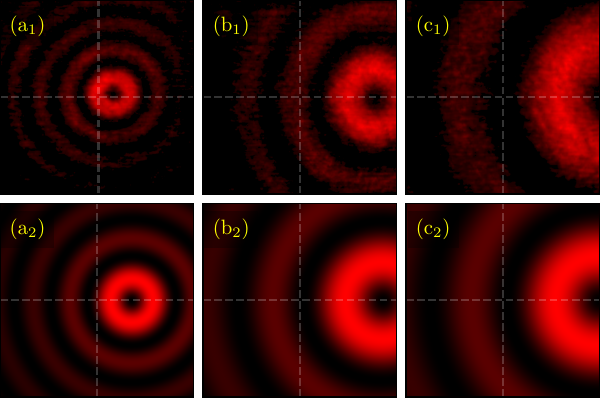}
    \caption{Experimental transverse intensity profiles of syncopated Bessel beams for $m=1$, measured at propagation distances of $z=10$ cm for $(\mathrm{a}_1)$ and $(\mathrm{a}_2)$, $z=20$ cm for $(\mathrm{b}_1)$ and $(\mathrm{b}_2)$, and $z=30$ cm for $(\mathrm{c}_1)$ and $(\mathrm{c}_2)$. In each pair, $(\mathrm{a}_1)$, $(\mathrm{b}_1)$, and $(\mathrm{c}_1)$ correspond to the experimental realizations, while $(\mathrm{a}_2)$, $(\mathrm{b}_2)$, and $(\mathrm{c}_2)$ are the analytical counterparts of these realizations. The fields were synthesized using a He-Ne laser ($\lambda=633$ nm) and a spatial light modulator (SLM) with an annular source of radius $r_0=120$ $\mu$m and modulation depth $u=4$~\cite{Arrizon:07}. The observation window is $2 \times 2$ mm, which is essentially the size of the beam. The white cross hairs indicate the optical axis ($x=y=0$), showing a clear radial displacement of the center of symmetry. This physical shift confirms the successful topological reconfiguration of the field, where the phase singularity of the $m=1$ mode remains structurally stable despite being displaced off-axis. The measured intensity distributions at different planes along the optical axis confirm that the beam maintains its self-scaling invariance, with the transverse profile expanding linearly while preserving its structural integrity.}
    \label{fig_3}
\end{figure}

\begin{figure}
\centering\includegraphics[width=\linewidth]{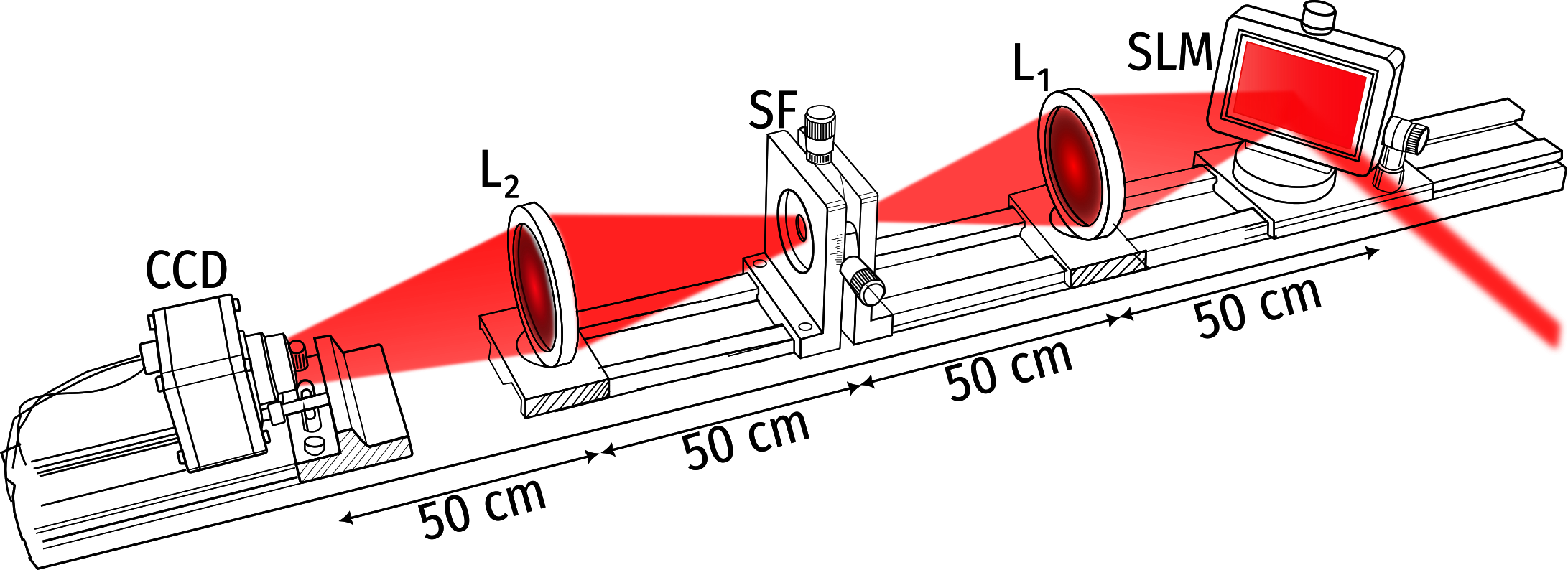}
    \caption{Schematic representation of the experimental setup for the synthesis and characterization of syncopated Bessel beams. A He-Ne laser ($\lambda = 633$ nm) illuminates a spatial light modulator (SLM) programmed with the syncopated phase profile via a pixelated complex-amplitude encoding protocol. A 4$f$ optical relay, comprising lenses $L_1$ and $L_2$ and a spatial filter (SF) positioned at the Fourier plane, is employed to isolate the desired diffraction order. The transverse intensity distributions were recorded at three distinct observation planes, as shown in Fig. \ref{fig_3}. These measurements serve to characterize the off-axis trajectory and experimentally verify the self-scaling invariance of the field.}
    \label{fig_4}
\end{figure}

The experimental synthesis of syncopated Bessel beams represents a fundamental departure from strict azimuthal symmetry towards tunable structural complexity. By encoding the syncopated phase profile onto a spatial light modulator via a pixelated complex-encoding protocol~\cite{Arrizon:07} (see Fig.~\ref{fig_4}), we observe a precise radial displacement that rigorously follows the analytical predictions of the syncopated coordinate $w$, as presented in Fig.~\ref{fig_3}. The measured intensity distributions reveal that while the beam's symmetry center is shifted relative to the optical axis, the fundamental self-scaling invariance of the mode is perfectly preserved. As evidenced by the volumetric visualizations in Fig.~\ref{fig_1}, the beam maintains its structural integrity despite the significant radial displacement. This shows that the sinusoidal phase rhythm establishes a stable off-axis propagation trajectory without structural degradation, allowing precise relocation of intensity maxima and phase vortex cores without the need for mechanical translation stages or complex optimization routines~\cite{roadmap2017}.

Analyzed through the Madelung-Bohm formalism, this experimental validation provides a profound physical insight. The transverse energy flow behaves exactly as dictated by the analytical potential $Q_{\texttt{Bohm}}$. The transverse phase-gradient pressure dynamically reconfigures the internal potential structure, forcing the asymmetric intensity distribution and its associated singularity into a stable off-axis trajectory. This physical mechanism perfectly explains the beam's resistance to diffractive decay, positioning the syncopated field as a highly resilient, structurally pristine mode for energy transport and topological manipulation.

The discovery and rigorous validation of syncopated Bessel beams establish a definitive regime in structured light, where a sinusoidal phase rhythm reconfigures the paraxial propagation dynamics through a stable structural guide. This analytical framework, anchored by the Graf addition theorem, reveals a robust protocol for spatial displacement without loss of topological stability. Our results confirm that this topological reconfiguration yields a self-scaling propagation mode that effectively steers optical singularities purely through electromagnetic phase control.

Ultimately, the realization of syncopated optical fields dismantles the conventional assumption that strict spatial symmetry is a prerequisite for diffraction-resilient propagation. By proving that spatial symmetry can be rigorously decoupled from structural invariance, this work introduces a fundamental topological degree of freedom in wave-field engineering. These findings catalyze a new era of agile beam steering, eliminating both the computational overhead of iterative holography and the latency of mechanical translation. The capacity to command stable, asymmetric potential wells off-axis opens unprecedented avenues for high-bandwidth orbital angular momentum multiplexing, advanced optical trapping, and high-precision metrology, cementing the syncopated coordinate transformation as a formidable foundation for next-generation structured light technologies.

\begin{acknowledgments}
I. Juli\'an-Mac\'ias thanks the Secretar\'ia de Ciencia, Humanidades, Tecnolog\'ia e Innovaci\'on (SECIHTI) for the postdoctoral fellowship awarded (No. CVU: 737102). The authors thank Lic. Adriana Elorza Villanueva from the technical committee at INAOE for the preparation of the schematic representation, Fig.~\ref{fig_4}.
\end{acknowledgments}


\begin{thebibliography}{18}%
\makeatletter
\providecommand \@ifxundefined [1]{%
 \@ifx{#1\undefined}
}%
\providecommand \@ifnum [1]{%
 \ifnum #1\expandafter \@firstoftwo
 \else \expandafter \@secondoftwo
 \fi
}%
\providecommand \@ifx [1]{%
 \ifx #1\expandafter \@firstoftwo
 \else \expandafter \@secondoftwo
 \fi
}%
\providecommand \natexlab [1]{#1}%
\providecommand \enquote  [1]{``#1''}%
\providecommand \bibnamefont  [1]{#1}%
\providecommand \bibfnamefont [1]{#1}%
\providecommand \citenamefont [1]{#1}%
\providecommand \href@noop [0]{\@secondoftwo}%
\providecommand \href [0]{\begingroup \@sanitize@url \@href}%
\providecommand \@href[1]{\@@startlink{#1}\@@href}%
\providecommand \@@href[1]{\endgroup#1\@@endlink}%
\providecommand \@sanitize@url [0]{\catcode `\\12\catcode `\$12\catcode `\&12\catcode `\#12\catcode `\^12\catcode `\_12\catcode `\%12\relax}%
\providecommand \@@startlink[1]{}%
\providecommand \@@endlink[0]{}%
\providecommand \url  [0]{\begingroup\@sanitize@url \@url }%
\providecommand \@url [1]{\endgroup\@href {#1}{\urlprefix }}%
\providecommand \urlprefix  [0]{URL }%
\providecommand \Eprint [0]{\href }%
\providecommand \doibase [0]{https://doi.org/}%
\providecommand \selectlanguage [0]{\@gobble}%
\providecommand \bibinfo  [0]{\@secondoftwo}%
\providecommand \bibfield  [0]{\@secondoftwo}%
\providecommand \translation [1]{[#1]}%
\providecommand \BibitemOpen [0]{}%
\providecommand \bibitemStop [0]{}%
\providecommand \bibitemNoStop [0]{.\EOS\space}%
\providecommand \EOS [0]{\spacefactor3000\relax}%
\providecommand \BibitemShut  [1]{\csname bibitem#1\endcsname}%
\let\auto@bib@innerbib\@empty
\bibitem [{\citenamefont {Durnin}(1987)}]{durnin1987}%
  \BibitemOpen
  \bibfield  {author} {\bibinfo {author} {\bibfnamefont {J.}~\bibnamefont {Durnin}},\ }\bibfield  {title} {\bibinfo {title} {Exact solutions for nondiffracting beams. {I}. {T}he scalar theory},\ }\href {https://doi.org/10.1364/JOSAA.4.000651} {\bibfield  {journal} {\bibinfo  {journal} {J. Opt. Soc. Am. A}\ }\textbf {\bibinfo {volume} {4}},\ \bibinfo {pages} {651} (\bibinfo {year} {1987})}\BibitemShut {NoStop}%
\bibitem [{\citenamefont {Bouchal}(2003)}]{bouchal2003}%
  \BibitemOpen
  \bibfield  {author} {\bibinfo {author} {\bibfnamefont {Z.}~\bibnamefont {Bouchal}},\ }\bibfield  {title} {\bibinfo {title} {Nondiffracting {O}ptical {B}eams: {P}hysical {P}roperties, {E}xperiments, and {A}pplications},\ }\href {https://doi.org/10.1023/A:1024802801048} {\bibfield  {journal} {\bibinfo  {journal} {Czechoslovak Journal of Physics}\ }\textbf {\bibinfo {volume} {53}},\ \bibinfo {pages} {537} (\bibinfo {year} {2003})}\BibitemShut {NoStop}%
\bibitem [{\citenamefont {McGloin}\ and\ \citenamefont {Dholakia}(2005)}]{mcgloin2005}%
  \BibitemOpen
  \bibfield  {author} {\bibinfo {author} {\bibfnamefont {D.}~\bibnamefont {McGloin}}\ and\ \bibinfo {author} {\bibfnamefont {K.}~\bibnamefont {Dholakia}},\ }\bibfield  {title} {\bibinfo {title} {Bessel beams: Diffraction in a new light},\ }\href {https://doi.org/10.1080/0010751042000275259} {\bibfield  {journal} {\bibinfo  {journal} {Contemporary Physics}\ }\textbf {\bibinfo {volume} {46}},\ \bibinfo {pages} {15} (\bibinfo {year} {2005})}\BibitemShut {NoStop}%
\bibitem [{\citenamefont {Rubinsztein-Dunlop}\ \emph {et~al.}(2016)\citenamefont {Rubinsztein-Dunlop}, \citenamefont {Forbes}, \citenamefont {Berry}, \citenamefont {Dennis}, \citenamefont {Andrews}, \citenamefont {Mansuripur}, \citenamefont {Denz}, \citenamefont {Alpmann}, \citenamefont {Banzer}, \citenamefont {Bauer}, \citenamefont {Karimi}, \citenamefont {Marrucci}, \citenamefont {Padgett}, \citenamefont {Ritsch-Marte}, \citenamefont {Litchinitser}, \citenamefont {Bigelow}, \citenamefont {Rosales-Guzmán}, \citenamefont {Belmonte}, \citenamefont {Torres}, \citenamefont {Neely}, \citenamefont {Baker}, \citenamefont {Gordon}, \citenamefont {Stilgoe}, \citenamefont {Romero}, \citenamefont {White}, \citenamefont {Fickler}, \citenamefont {Willner}, \citenamefont {Xie}, \citenamefont {McMorran},\ and\ \citenamefont {Weiner}}]{roadmap2017}%
  \BibitemOpen
  \bibfield  {author} {\bibinfo {author} {\bibfnamefont {H.}~\bibnamefont {Rubinsztein-Dunlop}}, \bibinfo {author} {\bibfnamefont {A.}~\bibnamefont {Forbes}}, \bibinfo {author} {\bibfnamefont {M.~V.}\ \bibnamefont {Berry}}, \bibinfo {author} {\bibfnamefont {M.~R.}\ \bibnamefont {Dennis}}, \bibinfo {author} {\bibfnamefont {D.~L.}\ \bibnamefont {Andrews}}, \bibinfo {author} {\bibfnamefont {M.}~\bibnamefont {Mansuripur}}, \bibinfo {author} {\bibfnamefont {C.}~\bibnamefont {Denz}}, \bibinfo {author} {\bibfnamefont {C.}~\bibnamefont {Alpmann}}, \bibinfo {author} {\bibfnamefont {P.}~\bibnamefont {Banzer}}, \bibinfo {author} {\bibfnamefont {T.}~\bibnamefont {Bauer}}, \bibinfo {author} {\bibfnamefont {E.}~\bibnamefont {Karimi}}, \bibinfo {author} {\bibfnamefont {L.}~\bibnamefont {Marrucci}}, \bibinfo {author} {\bibfnamefont {M.}~\bibnamefont {Padgett}}, \bibinfo {author} {\bibfnamefont {M.}~\bibnamefont {Ritsch-Marte}}, \bibinfo {author} {\bibfnamefont {N.~M.}\ \bibnamefont {Litchinitser}}, \bibinfo {author} {\bibfnamefont {N.~P.}\ \bibnamefont {Bigelow}}, \bibinfo {author} {\bibfnamefont {C.}~\bibnamefont {Rosales-Guzmán}}, \bibinfo {author} {\bibfnamefont {A.}~\bibnamefont {Belmonte}}, \bibinfo {author} {\bibfnamefont {J.~P.}\ \bibnamefont {Torres}}, \bibinfo {author} {\bibfnamefont {T.~W.}\ \bibnamefont {Neely}}, \bibinfo {author} {\bibfnamefont {M.}~\bibnamefont {Baker}}, \bibinfo {author} {\bibfnamefont {R.}~\bibnamefont {Gordon}}, \bibinfo {author} {\bibfnamefont {A.~B.}\ \bibnamefont {Stilgoe}}, \bibinfo {author} {\bibfnamefont {J.}~\bibnamefont {Romero}}, \bibinfo {author} {\bibfnamefont {A.~G.}\ \bibnamefont {White}}, \bibinfo {author} {\bibfnamefont {R.}~\bibnamefont {Fickler}}, \bibinfo {author} {\bibfnamefont {A.~E.}\ \bibnamefont {Willner}}, \bibinfo {author} {\bibfnamefont {G.}~\bibnamefont {Xie}}, \bibinfo {author} {\bibfnamefont {B.}~\bibnamefont {McMorran}},\ and\ \bibinfo {author} {\bibfnamefont {A.~M.}\ \bibnamefont {Weiner}},\ }\bibfield  {title} {\bibinfo {title} {Roadmap on structured light},\ }\href {https://doi.org/10.1088/2040-8978/19/1/013001} {\bibfield  {journal} {\bibinfo  {journal} {Journal of Optics}\ }\textbf {\bibinfo {volume} {19}},\ \bibinfo {pages} {013001} (\bibinfo {year} {2016})}\BibitemShut {NoStop}%
\bibitem [{\citenamefont {Gori}\ \emph {et~al.}(1987)\citenamefont {Gori}, \citenamefont {Guattari},\ and\ \citenamefont {Padovani}}]{gori1987}%
  \BibitemOpen
  \bibfield  {author} {\bibinfo {author} {\bibfnamefont {F.}~\bibnamefont {Gori}}, \bibinfo {author} {\bibfnamefont {G.}~\bibnamefont {Guattari}},\ and\ \bibinfo {author} {\bibfnamefont {C.}~\bibnamefont {Padovani}},\ }\bibfield  {title} {\bibinfo {title} {Bessel-{G}auss beams},\ }\href {https://doi.org/https://doi.org/10.1016/0030-4018(87)90276-8} {\bibfield  {journal} {\bibinfo  {journal} {Optics Communications}\ }\textbf {\bibinfo {volume} {64}},\ \bibinfo {pages} {491} (\bibinfo {year} {1987})}\BibitemShut {NoStop}%
\bibitem [{\citenamefont {Berry}\ and\ \citenamefont {Balazs}(1979)}]{berry1979}%
  \BibitemOpen
  \bibfield  {author} {\bibinfo {author} {\bibfnamefont {M.~V.}\ \bibnamefont {Berry}}\ and\ \bibinfo {author} {\bibfnamefont {N.~L.}\ \bibnamefont {Balazs}},\ }\bibfield  {title} {\bibinfo {title} {Nonspreading wave packets},\ }\href {https://doi.org/10.1119/1.11855} {\bibfield  {journal} {\bibinfo  {journal} {American Journal of Physics}\ }\textbf {\bibinfo {volume} {47}},\ \bibinfo {pages} {264} (\bibinfo {year} {1979})}\BibitemShut {NoStop}%
\bibitem [{\citenamefont {Siviloglou}\ \emph {et~al.}(2007)\citenamefont {Siviloglou}, \citenamefont {Broky}, \citenamefont {Dogariu},\ and\ \citenamefont {Christodoulides}}]{siviloglou2007}%
  \BibitemOpen
  \bibfield  {author} {\bibinfo {author} {\bibfnamefont {G.~A.}\ \bibnamefont {Siviloglou}}, \bibinfo {author} {\bibfnamefont {J.}~\bibnamefont {Broky}}, \bibinfo {author} {\bibfnamefont {A.}~\bibnamefont {Dogariu}},\ and\ \bibinfo {author} {\bibfnamefont {D.~N.}\ \bibnamefont {Christodoulides}},\ }\bibfield  {title} {\bibinfo {title} {Observation of {A}ccelerating {A}iry {B}eams},\ }\href {https://doi.org/10.1103/PhysRevLett.99.213901} {\bibfield  {journal} {\bibinfo  {journal} {Phys. Rev. Lett.}\ }\textbf {\bibinfo {volume} {99}},\ \bibinfo {pages} {213901} (\bibinfo {year} {2007})}\BibitemShut {NoStop}%
\bibitem [{\citenamefont {Kotlyar}\ \emph {et~al.}(2014)\citenamefont {Kotlyar}, \citenamefont {Kovalev}, \citenamefont {Skidanov},\ and\ \citenamefont {Soifer}}]{kotlyar2014}%
  \BibitemOpen
  \bibfield  {author} {\bibinfo {author} {\bibfnamefont {V.~V.}\ \bibnamefont {Kotlyar}}, \bibinfo {author} {\bibfnamefont {A.~A.}\ \bibnamefont {Kovalev}}, \bibinfo {author} {\bibfnamefont {R.~V.}\ \bibnamefont {Skidanov}},\ and\ \bibinfo {author} {\bibfnamefont {V.~A.}\ \bibnamefont {Soifer}},\ }\bibfield  {title} {\bibinfo {title} {Asymmetric {B}essel-{G}auss beams},\ }\href {https://doi.org/10.1364/JOSAA.31.001977} {\bibfield  {journal} {\bibinfo  {journal} {J. Opt. Soc. Am. A}\ }\textbf {\bibinfo {volume} {31}},\ \bibinfo {pages} {1977} (\bibinfo {year} {2014})}\BibitemShut {NoStop}%
\bibitem [{\citenamefont {Abramowitz}\ and\ \citenamefont {Stegun}(1964)}]{abramowitz1964handbook}%
  \BibitemOpen
  \bibfield  {author} {\bibinfo {author} {\bibfnamefont {M.}~\bibnamefont {Abramowitz}}\ and\ \bibinfo {author} {\bibfnamefont {I.~A.}\ \bibnamefont {Stegun}},\ }\href@noop {} {\emph {\bibinfo {title} {Handbook of Mathematical Functions with Formulas, Graphs, and Mathematical Tables}}},\ \bibinfo {series} {Applied Mathematics Series}\ No.~\bibinfo {number} {55}\ (\bibinfo  {publisher} {National Bureau of Standards},\ \bibinfo {address} {Washington, DC},\ \bibinfo {year} {1964})\BibitemShut {NoStop}%
\bibitem [{\citenamefont {Colton}\ and\ \citenamefont {Kress}(1998)}]{Colton1998}%
  \BibitemOpen
  \bibfield  {author} {\bibinfo {author} {\bibfnamefont {D.}~\bibnamefont {Colton}}\ and\ \bibinfo {author} {\bibfnamefont {R.}~\bibnamefont {Kress}},\ }\href@noop {} {\emph {\bibinfo {title} {Inverse acoustic and electromagnetic scattering theory}}},\ \bibinfo {edition} {2nd}\ ed.,\ \bibinfo {series} {Applied mathematical sciences}\ No.~\bibinfo {number} {93}\ (\bibinfo  {publisher} {Springer},\ \bibinfo {address} {Berlin},\ \bibinfo {year} {1998})\ \bibinfo {note} {literaturverz. S. [318] - 331}\BibitemShut {NoStop}%
\bibitem [{\citenamefont {Cuyt}(2008)}]{Cuyt2008}%
  \BibitemOpen
  \bibfield  {author} {\bibinfo {author} {\bibfnamefont {A.}~\bibnamefont {Cuyt}},\ }\href@noop {} {\emph {\bibinfo {title} {Handbook of Continued Fractions for Special Functions}}},\ edited by\ \bibinfo {editor} {\bibfnamefont {F.}~\bibnamefont {Backeljauw}}, \bibinfo {editor} {\bibfnamefont {C.}~\bibnamefont {Bonan-Hamada}}, \bibinfo {editor} {\bibfnamefont {V.}~\bibnamefont {Petersen}}, \bibinfo {editor} {\bibfnamefont {B.}~\bibnamefont {Verdonk}}, \bibinfo {editor} {\bibfnamefont {H.}~\bibnamefont {Waadeland}},\ and\ \bibinfo {editor} {\bibfnamefont {W.~B.}\ \bibnamefont {Jones}}\ (\bibinfo  {publisher} {Springer Netherlands},\ \bibinfo {address} {Dordrecht},\ \bibinfo {year} {2008})\ \bibinfo {note} {description based on publisher supplied metadata and other sources.}\BibitemShut {Stop}%
\bibitem [{\citenamefont {Olver}\ \emph {et~al.}(2010)\citenamefont {Olver}, \citenamefont {Lozier}, \citenamefont {Boisvert},\ and\ \citenamefont {Clark}}]{Olver2010}%
  \BibitemOpen
  \bibinfo {editor} {\bibfnamefont {F.~W.~J.}\ \bibnamefont {Olver}}, \bibinfo {editor} {\bibfnamefont {D.~W.}\ \bibnamefont {Lozier}}, \bibinfo {editor} {\bibfnamefont {R.~F.}\ \bibnamefont {Boisvert}},\ and\ \bibinfo {editor} {\bibfnamefont {C.~W.}\ \bibnamefont {Clark}},\ eds.,\ \href@noop {} {\emph {\bibinfo {title} {NIST handbook of mathematical functions}}}\ (\bibinfo  {publisher} {Cambridge University Press},\ \bibinfo {address} {Cambridge},\ \bibinfo {year} {2010})\ \bibinfo {note} {neubearbeitung von: Handbook of mathematical functions with formulas, graphs, and mathematical tables / M. Abramowitz and I.A. Stegun, editors (1964)}\BibitemShut {NoStop}%
\bibitem [{\citenamefont {Dattoli}\ \emph {et~al.}(1990)\citenamefont {Dattoli}, \citenamefont {Giannessi}, \citenamefont {Mezi},\ and\ \citenamefont {Torre}}]{Dattoli1990}%
  \BibitemOpen
  \bibfield  {author} {\bibinfo {author} {\bibfnamefont {G.}~\bibnamefont {Dattoli}}, \bibinfo {author} {\bibfnamefont {L.}~\bibnamefont {Giannessi}}, \bibinfo {author} {\bibfnamefont {L.}~\bibnamefont {Mezi}},\ and\ \bibinfo {author} {\bibfnamefont {A.}~\bibnamefont {Torre}},\ }\bibfield  {title} {\bibinfo {title} {Theory of generalized {B}essel functions},\ }\href {https://doi.org/10.1007/BF02726105} {\bibfield  {journal} {\bibinfo  {journal} {Il Nuovo Cimento B (1971-1996)}\ }\textbf {\bibinfo {volume} {105}},\ \bibinfo {pages} {327} (\bibinfo {year} {1990})}\BibitemShut {NoStop}%
\bibitem [{\citenamefont {Madelung}(1927)}]{Madelung_1927}%
  \BibitemOpen
  \bibfield  {author} {\bibinfo {author} {\bibfnamefont {E.}~\bibnamefont {Madelung}},\ }\bibfield  {title} {\bibinfo {title} {Quantentheorie in hydrodynamischer {F}orm},\ }\href {https://doi.org/10.1007/BF01400372} {\bibfield  {journal} {\bibinfo  {journal} {Zeitschrift f{\"u}r Physik}\ }\textbf {\bibinfo {volume} {40}},\ \bibinfo {pages} {322} (\bibinfo {year} {1927})}\BibitemShut {NoStop}%
\bibitem [{\citenamefont {Bohm}(1952)}]{Bohm_1952}%
  \BibitemOpen
  \bibfield  {author} {\bibinfo {author} {\bibfnamefont {D.}~\bibnamefont {Bohm}},\ }\bibfield  {title} {\bibinfo {title} {A {S}uggested {I}nterpretation of the {Q}uantum {T}heory in {T}erms of ``{H}iden'' {V}ariables. {I}},\ }\href {https://doi.org/10.1103/PhysRev.85.166} {\bibfield  {journal} {\bibinfo  {journal} {Phys. Rev.}\ }\textbf {\bibinfo {volume} {85}},\ \bibinfo {pages} {166} (\bibinfo {year} {1952})}\BibitemShut {NoStop}%
\bibitem [{\citenamefont {Hojman}\ \emph {et~al.}(2021)\citenamefont {Hojman}, \citenamefont {Asenjo}, \citenamefont {Moya-Cessa},\ and\ \citenamefont {Soto-Eguibar}}]{Hojman2021}%
  \BibitemOpen
  \bibfield  {author} {\bibinfo {author} {\bibfnamefont {S.~A.}\ \bibnamefont {Hojman}}, \bibinfo {author} {\bibfnamefont {F.~A.}\ \bibnamefont {Asenjo}}, \bibinfo {author} {\bibfnamefont {H.~M.}\ \bibnamefont {Moya-Cessa}},\ and\ \bibinfo {author} {\bibfnamefont {F.}~\bibnamefont {Soto-Eguibar}},\ }\bibfield  {title} {\bibinfo {title} {Bohm potential is real and its effects are measurable},\ }\href {https://doi.org/https://doi.org/10.1016/j.ijleo.2021.166341} {\bibfield  {journal} {\bibinfo  {journal} {Optik}\ }\textbf {\bibinfo {volume} {232}},\ \bibinfo {pages} {166341} (\bibinfo {year} {2021})}\BibitemShut {NoStop}%
\bibitem [{\citenamefont {Moya-Cessa}\ \emph {et~al.}(2024)\citenamefont {Moya-Cessa}, \citenamefont {Ramos-Prieto}, \citenamefont {{S\'anchez}-{de-la-Llave}}, \citenamefont {Ru\'{\i}z}, \citenamefont {Arriz\'on},\ and\ \citenamefont {Soto-Eguibar}}]{CRB}%
  \BibitemOpen
  \bibfield  {author} {\bibinfo {author} {\bibfnamefont {H.~M.}\ \bibnamefont {Moya-Cessa}}, \bibinfo {author} {\bibfnamefont {I.}~\bibnamefont {Ramos-Prieto}}, \bibinfo {author} {\bibfnamefont {D.}~\bibnamefont {{S\'anchez}-{de-la-Llave}}}, \bibinfo {author} {\bibfnamefont {U.}~\bibnamefont {Ru\'{\i}z}}, \bibinfo {author} {\bibfnamefont {V.}~\bibnamefont {Arriz\'on}},\ and\ \bibinfo {author} {\bibfnamefont {F.}~\bibnamefont {Soto-Eguibar}},\ }\bibfield  {title} {\bibinfo {title} {Cauchy-{R}iemann beams},\ }\href {https://doi.org/10.1103/PhysRevA.109.043528} {\bibfield  {journal} {\bibinfo  {journal} {Phys. Rev. A}\ }\textbf {\bibinfo {volume} {109}},\ \bibinfo {pages} {043528} (\bibinfo {year} {2024})}\BibitemShut {NoStop}%
\bibitem [{\citenamefont {Arriz\'{o}n}\ \emph {et~al.}(2007)\citenamefont {Arriz\'{o}n}, \citenamefont {Ruiz}, \citenamefont {Carrada},\ and\ \citenamefont {Gonz\'{a}lez}}]{Arrizon:07}%
  \BibitemOpen
  \bibfield  {author} {\bibinfo {author} {\bibfnamefont {V.}~\bibnamefont {Arriz\'{o}n}}, \bibinfo {author} {\bibfnamefont {U.}~\bibnamefont {Ruiz}}, \bibinfo {author} {\bibfnamefont {R.}~\bibnamefont {Carrada}},\ and\ \bibinfo {author} {\bibfnamefont {L.~A.}\ \bibnamefont {Gonz\'{a}lez}},\ }\bibfield  {title} {\bibinfo {title} {Pixelated phase computer holograms for the accurate encoding of scalar complex fields},\ }\href {https://doi.org/10.1364/JOSAA.24.003500} {\bibfield  {journal} {\bibinfo  {journal} {J. Opt. Soc. Am. A}\ }\textbf {\bibinfo {volume} {24}},\ \bibinfo {pages} {3500} (\bibinfo {year} {2007})}\BibitemShut {NoStop}%
\end{thebibliography}
%

\end{document}